\RequirePackage{fix-cm}
\documentclass[smallextended]{svjour3}       
\usepackage{graphicx}
\usepackage{enumerate}
\usepackage[misc]{ifsym}
\usepackage{color}

\begin{document}

\title{ An Improved Quantum Private Set Intersection Protocol Based on Hadamard Gates 
}
\author{Wen-Jie Liu \textsuperscript{1,2}        \and
        Wen-Bo Li \textsuperscript{1}       \and
        Hai-Bin Wang\textsuperscript{1,2}
}


\institute{%
    \begin{itemize}
      \item[\textsuperscript{\Letter}] {Wen-Jie Liu} \\
            {wenjiel@163.com}
       \item[]{Wen-Bo Li} \\
            {lwb0408@163.com}
      \at
      \item[\textsuperscript{1}] School of Computer and Software, Nanjing University of Information Science and Technology, Nanjing 210044, China
      \item[\textsuperscript{2}] 
      Engineering Research Center of Digital Forensics, Ministry of Education, Nanjing 210044, China
    \end{itemize}
}

\date{Received: date / Accepted: date}

\maketitle

\begin{abstract}
Recently, Liu and Yin (Int. J. Theor. Phys. 60, 2074-2083 (2021)) proposed a two-party private set intersection protocol based on quantum Fourier transform. We find the participant can deduce the other party’s private information, which  violates the security requirement of private set computation. In order to solve this problem, an improved  private set intersection protocol based on Hadamard gate is proposed. Firstly, the more feasible Hadamard gates are used to perform on the original $n$ qubits instead of the quantum Fourier transform, which may reduce the difficulty of implementation. In addition, through the exclusive OR calculation, the participant’s private information is randomly chosen and encoded on the additional  $n$ qubits, which prevents participants from obtaining the result of the difference set $S_{diff}$, and then avoids  the internal leakage of private information. Finally, the correctness and security analysis are conducted to show the proposed protocol  can guarantee the correctness of computation result as well as resist outside attacks and participant internal attacks.


\end{abstract}

\section{Introduction}
\label{intro:1}
Secure multiparty computation (SMC) is a collaborative computing problem that derived from the ``Millionaire'' problem \cite{RefY82} raised by Yao in 1982.  Under the premise of correct computation, the private information of participants who do not trust each other will not be leaked. Private set computation(PSC) is an important aspect of SMC. It is the computing foundation of data mining and machine learning based on privacy protection, and it is also one of the active research topics in the classic information security field in recent years.
At present, the security of classical PSC protocol is basically based on computational complexity, and its security can be guaranteed under the condition of limited computing power. 
However, quantum computing has shown super-parallel computing capabilities that classical computing cannot match, such as solving RSA large prime factorization problem \cite{RefS94}, secondary acceleration of out-of-order database retrieval \cite{RefG97}, and quantum forgery attacks on COPA, AES-COPA and marble authenticated encryption algorithms \cite{RefY21} etc. The security of most classical protocols is not guaranteed in this situation. In this regard, many scholars have begun to study quantum algorithm \cite{RefL18,RefL19}. The privacy and security of quantum algorithm is based on the physical properties of quantum mechanics, such as the No-Cloning theorem \cite{RefW82}, uncertainty principle \cite{RefF97}, quantum entanglement, potential unconditional security \cite{RefM98}, etc. At present, the corresponding study of QPSC mainly includes quantum private set intersection (QPSI) \cite{RefS16,RefC16,RefM18,RefD21,RefL21} and quantum private set cardinality \cite{RefS18,RefS182,RefS19,RefL20}.

Private set intersection is an important cryptographic primitive that performs joint operations on data sets in a privacy-protected manner. Specifically, multiple participants calculate the intersection without revealing their privacy to others (including internal participants). In 2016, Shi et al. \cite{RefS16} first proposed a quantum scheme for the intersection of private sets. However, the server could unilaterally manipulate the intersection results in this protocol. In order to solve this problem, Cheng et al. \cite{RefC16} introduced a passive third party to achieve the fairness of the protocol.
After that, based on the decision-making protocol for oblivious set members \cite{RefS15}, Maitra \cite{RefM18} proposed quantum secure two-party computation for set intersection with rational players. 
These studies \cite{RefS16,RefC16,RefM18} require ``multi-particle entangled states" as quantum resources and some ``complex oracle operators", which are difficult to achieve under current technology. 
In 2021, Debnath \cite{RefD21} proposed a QPSI protocol between the client and the server, which has higher feasibility using single photon quantum resources.
But the result of the intersection in this protocol can only be obtained by one participant (i.e., client).
Recently, Liu et al. proposed a novel QPSI protocol \cite{RefL21} (we call it the NQPSI protocol) based on quantum Fourier transform. According to the published results, all participants can obtain the intersection. However, private information will be leaked inside the participants. 
In order to solve this problem, an improved QPSI protocol based on Hadamard ($\rm H$) gate is proposed. This protocol has two obvious advantages. Firstly, the more feasible $\rm H$ gate is used to replace the original quantum Fourier transform, which reduces the difficulty of implementing the protocol. More importantly, through exclusive OR calculation, 
the participant’s private information is randomly chosen and encoded on the additional  $n$ qubits, which prevents participants from getting the result of difference set, and then avoids  the internal leakage of private information.

The rest of the paper is organized as follows. Sect. 2 reviews the NQPSI protocol and analyzes its loopholes. In Sect. 3, an improved private set intersection protocol is proposed. Sect. 4 verifies the correctness of the two protocols through examples and analyzes the security when facing outside attacks and participant attacks. Sect. 5 gives a brief summary of the content of this paper and prospects for future work.
\section{Review and analysis of NQPSI protocol}
For clarity, the NQPSI protocol \cite{RefL21} of Liu et al. is reviewed and analyzed here. The specific content is as follows.
\subsection{Review on the NQPSI protocol}
First, they deduced that the second quantum Fourier transform of a single qubit is $QF{T^2}\left| j \right\rangle  = -\left| j \right\rangle $, then  $QF{T^4}\left| j \right\rangle  = \left| j \right\rangle $.
Suppose there is a complete set $U = \left\{ {{x_1},{x_2}, \cdots {x_n}} \right\}$. 
Participants Alice and Bob have private sets ${S_A} = \left\{ {s_1^A,s_2^A, \cdots s_{{l_A}}^A} \right\}$ and ${S_B} = \left\{ {s_1^B,s_2^B, \cdots s_{{l_A}}^B} \right\}$ respectively, where ${S_A},{S_B} \subseteq U$. Alice and Bob coded their private set codes as ${C_A} = \left\{ {c_{_1}^A,c_{_2}^A, \cdots c_n^A} \right\}$ and ${C_B} = \left\{ {c_{_1}^B,c_{_2}^B, \cdots c_n^B} \right\}$ respectively. The coding rules are shown in Eq.~(\ref{eq1})  .
  \begin{equation}
\label{eq1}
c_i^A = \left\{
\begin{array}{rcl}
{1,if} & {{x_i} \in {S_A}}\\
{0,if} & {{x_i} \notin {S_A}}
\end{array} \right.,\;c_i^B = \left\{
\begin{array}{rcl}
{1,if} & {{x_i} \in {S_B}}\\
{0,if} & {{x_i} \notin {S_B}}
\end{array} \right..
\end{equation}
Calvin is a semi-honest third party. 
The steps of their protocol are as follows.
 \begin{enumerate}[Step 1]
\item Calvin prepares a particles sequence ${P_C} = \left\{ {p_1^C,p_2^C, \cdots ,p_n^C} \right\}$. He inserts the decoy photon into ${P_C}$ to form a new quantum sequence and sent it to Alice.
\item Alice verifies the decoy particles. If the verification result is correct, she discards the decoy photon and continues the next step. Otherwise, the protocol will be aborted.
\item Alice prepares two $n$-length strings ${R_A} = \left\{ {r_1^A,r_2^A, \cdots ,r_n^A} \right\}$ (The subscript $n$ here is written as $n+l$ in the original protocol, which is probably a typing error of the author) and ${H_A} = \left\{ {h_1^A,h_2^A, \cdots ,h_n^A} \right\}$, where $r_i^A\left( {i = 1, \cdots ,n} \right)$ is randomly chosen from $\left\{ {0,1} \right\}$ and $h_i^A\left( {i = 1, \cdots ,n} \right)$ is a random positive integer. Then she gets the quantum sequence ${P_A} = \left\{ {p_1^A,p_2^A, \cdots ,p_n^A} \right\}$, where $p_i^A = QF{T^{c_i^A \times 2}}QF{T^{r_i^A \times h_i^A}}p_i^C\left( {i = 1, \cdots ,n} \right)$. She inserts the decoy photon into ${P_A}$ to form a new quantum sequence and sends it to Bob.
\item Bob verifies the decoy particles. If the verification result is correct, he discards the decoy photon and continues the next step. Otherwise, the protocol will be  aborted.
\item  Bob prepares two $n$-length strings ${R_B}$ and ${H_B}$, and he gets the quantum sequence ${P_B} = \left\{ {p_1^B,p_2^B, \cdots ,p_n^B} \right\}$, where $p_i^B = QF{T^{c_i^B \times 2}}QF{T^{r_i^B \times h_i^B}}p_i^A$ $\left( {i = 1, \cdots ,n} \right)$. The rules are similar to Step 3. He inserts the decoy photon into ${P_A}$ to form a new quantum sequence and sends it to  Calvin.
\item Calvin verifies the decoy particles. If the verification result is correct, he discards the decoy photon and continues the next step. Otherwise, the protocol will be  aborted.
\item Alice and Bob compute $h_i^C = 4 -  {\left( {\left( {r_i^A \times h_i^A} \right) + \left( {r_i^B \times h_i^B} \right)} \right)\bmod 4} \left( {i = 1, \cdots ,n} \right)$ and send $h_1^C, \cdots ,h_n^C$ to Calvin. Calvin calculates $p_i^{C'} = QF{T^{h_i^C}}p_i^B$ and measures it. If the measurement result of $p_i^{C'}$ is the same as $p_i^C$, Calvin knows whether $c_i^A$ and $c_i^B$ are equal. Calvin gets ${S_A} \cap {S_B}$.
\end{enumerate}
 
\begin{table}
\caption{Results of the NQPSI protocol}
\label{tab:2}     
\begin{tabular}{llll}
\hline\noalign{\smallskip}
 Case &  $c_i^A $ & $c_i^B$ & $p_i^{C'}$ \\
\noalign{\smallskip}\hline\noalign{\smallskip}
1 & 0 & 0 & $p_i^C$ \\
2 & 0 & 1 & $-p_i^C$ \\ 
3 & 1 & 0 & $-p_i^C$ \\
4 & 1 & 1 & $ p_i^C$ \\
\noalign{\smallskip}\hline
\end{tabular}
\end{table}
\subsection{Vulnerability analysis}
In Step 7, Calvin can get 
\begin{equation}
p_i^{C'} = QF{T^{h_i^C}}p_i^B = QF{T^{c_i^A \times 2 + c_i^B \times 2}}p_i^C.
\end{equation}
Table~\ref{tab:2} shows the value of $p_i^{C'}$ for several paired $c_i^A$ and $c_i^B$. Each of the four different pairs ($c_i^A$, $c_i^B$) corresponds to results: $p_i^{C'}=p_i^C$ or $p_i^{C'}=-p_i^C$. Specifically, as shown in Case 1 and Case 4, $p_i^{C'}=p_i^C$ is the measurement result corresponding to the  ``complement-intersection'' set, ${{S}_{c-in}}={{S}_{com}}\cup{S_{in}}={{\cal C}_U}\left( {{S_A} \cup {S_B}} \right)\cup({S_A} \cap {S_B})$. In Case 2 and Case 3, $p_i^{C'}=-p_i^C$ is the result of difference set, ${S_{diff}} = \left( {{S_A} - {S_B}} \right) \cup \left( {{S_B} - {S_A}} \right)$.

Alice and Bob can obtain ${S_{in}}$ according to whether the $i$-th information corresponding to $p_i^{C'}=p_i^C$ is included in their private set or not.
However, they can know the $i$-th information corresponding to $p_i^{C'}=-p_i^C$ only contained in the own or the others private set. 
In Case 2, Alice can deduce that the $i$-th information is in the ${S_B}$. In Case 3, Bob can deduce that the $i$-th information is in the ${S_A}$. The privacy of set intersection requires that Alice (Bob) cannot know Bob's (Alice's) private set elements outside ${S_{in}}$. Obviously, the privacy of the NQPSI protocol cannot be guaranteed.

Furthermore, the author claims that Calvin can get ${S_{in}}$, which is impossible according to the protocol. And the author did not write the operation after Calvin got $c_i^A  = c_i^B$ in the protocol.

\section{QPSI protocol based on $\rm H$ gates}
\begin{figure}
\includegraphics[width=3in]{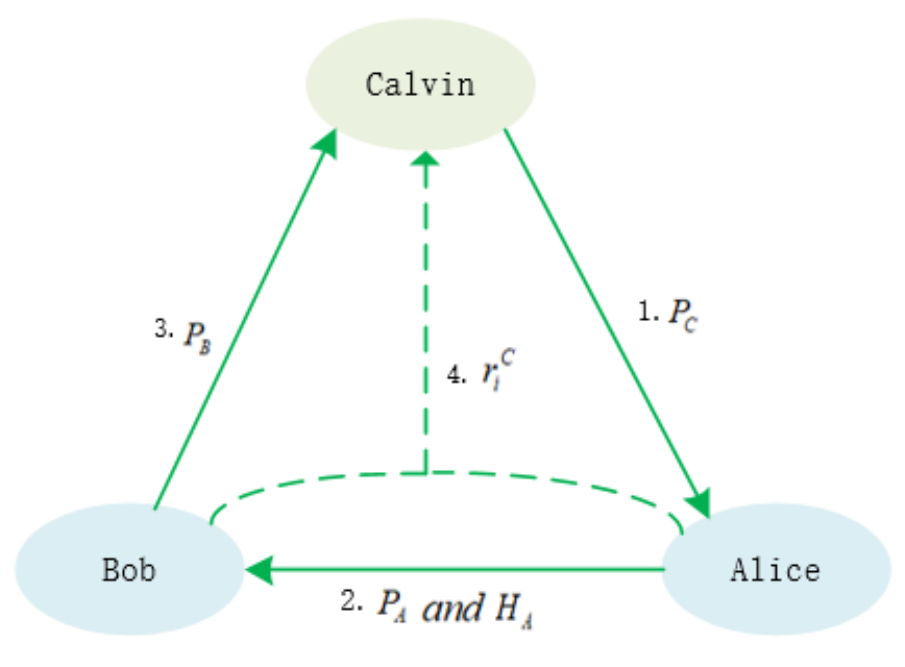}
\caption{Schematic diagram of QPSI protocol based on $\rm H$ gates. The line is the process from Step 1 to Step 6 in the protocol. Insertion of decoy photons and the eavesdropper detection are omitted here.}
\label{fig:1}   
\end{figure}

From the analysis in the previous section, participants can obtain private information outside ${S_{in}}$ of others based on ${S_{diff}}$, which leads to privacy leakage. 
In order to solve this problem, an improved QPSI protocol based on $\rm H$ gates is proposed, where ${\rm H} \equiv {1 \over {\sqrt 2 }}\left( {\matrix{
   1 & 1  \cr 
   1 & { - 1}  \cr 
 } } \right)$. The detailed procedures of our protocol are as follows (also shown in Fig.~\ref{fig:1}).
\begin{enumerate}[Step 1]
\item Calvin prepares the quantum sequence ${P_{C}} = \mathop  \otimes  \limits_{i = 1}^{2n} P_i^{C}$, where $P_i^{C}$ is randomly selected from $\left\{ {\left| 0 \right\rangle ,\left| 1 \right\rangle ,\left|  +  \right\rangle ,\left|  -  \right\rangle } \right\}$. Then he executes the operation of inserting decoy particle. Specifically, he prepares ${l_{C}}$ decoy particles $q_j^{C}$ and selects the measurement basis. After inserting $q_j^{C}$ into ${P_{C}}$ to form a new sequence ${P'_{C}}$, Calvin records the position $m_t^{C}\left( {t = 1,2, \cdots {l_{C}}} \right)$ of $q_j^{C}$ in sequence ${P'_{C}}$ and sends sequence ${P'_{C}}$ to Alice.

\item After receiving the sequence ${P'_{C}}$, Alice executes the eavesdropper detection. To be specific, Calvin sends the position $m_t^{C}$ of the decoy particle $q_j^{C}$ in the sequence ${P'_{C}}$ and the corresponding measurement basis to Alice. Alice compares the measurement result at this position with $q_j^{C}$. If ${q_j^{C}}^\prime$ is different from $q_j^{C}$, return to Step 1. Otherwise, she discards $q_j^{C}$ and proceeds to the next step.

\item Alice prepares $2n$-length strings ${R_{A}} = \left\{{r_i^{A}\left| r_i^{A} \in {{\rm N}^ + }, {i = 1,2, \cdots 2n} \right. } \right\}$ and ${\rm H_A} = \left\{ {h_i^A\left| {h_i^A \in \left\{ {0,1} \right\},n < i \le 2n} \right.} \right\}$. And she gets
\begin{equation}
\label{eq3}
\begin{array}{rcl} 
{P_A} &&= \mathop  \otimes \limits_{i = 1}^{2n} P_i^A \\
&&= \mathop  \otimes \limits_{i = 1}^n {{\rm H}^{{1 \over 3}\left( {c_i^A + 2} \right)}}{{\rm H}^{r_i^A}}P_i^C\left\| {\mathop  \otimes \limits_{i = n + 1}^{2n} {{\rm H}^{{1 \over 3}\left( {h_i^Ac_i^A + 2} \right)}}{{\rm H}^{r_i^A}}} \right.,
\end{array}
\end{equation}
where ${c_i^{A}}={c_{i-n}^{A}}$ ($i > n$). Then she executes the operation of inserting decoy particles, which is similar to Step 1. After this, she sends sequence ${P'_{A}}$ and  ${\rm H_A}$ to Bob.

\item After receiving sequence ${P'_{A}}$, Bob executes the eavesdropper detection, which is similar to Step 2.

\item Bob prepares $2n$-length strings ${R_{B}} = \left\{ {r_i^{B}\left| r_i^{B} \in {{\rm N}^ + }, {i = 1,2, \cdots 2n} \right.} \right\}$ and ${\rm H_B} = \left\{ {h_i^B\left| {h_i^B = h_i^A \oplus 1,n < i \le 2n} \right.} \right\}$. And he gets
\begin{equation}
\label{eq4}
\begin{array}{rcl} 
{P_B} &&= \mathop  \otimes \limits_{i = 1}^{2n} P_i^B \\
&&= \mathop  \otimes \limits_{i = 1}^n {{\rm H}^{{1 \over 3}\left( {c_i^B + 2} \right)}}{{\rm H}^{r_i^B}}P_i^A\left\| {\mathop  \otimes \limits_{i = n + 1}^{2n} {{\rm H}^{{1 \over 3}\left( {h_i^Bc_i^B + 2} \right)}}{{\rm H}^{r_i^B}}P_i^A} \right.,
\end{array}
\end{equation}
where ${c_i^{B}}={c_{i-n}^{B}}$ ($i > n$). Then he executes the operation of inserting decoy particles, which is similar to Step 1. After this, he sends sequence ${P'_{B}}$ to Calvin.

\item After receiving sequence ${P'_{B}}$, Calvin executes the eavesdropper detection, which is similar to Step 2.
\item Alice and Bob calculate $r_i^{C} = r_i^{A} + r_i^{B}$ and send the result to Calvin. Calvin gets 
\begin{equation}
\label{eq5}
\begin{array}{rcl} 
{P_{C}}^{\prime \prime } &&= \mathop \otimes \limits_{i = 1}^{2n} {P_i}^{{C}^{\prime \prime }} \\
&&= \mathop \otimes \limits_{i = 1}^{2n}{{\rm H}^{r_i^{C}}}P_i^{B} \\
&&= \mathop  \otimes \limits_{i = 1}^n {{\rm H}^{{1 \over 3}\left( {c_i^B + c_i^A + 4} \right)}}P_i^C\left\| {\mathop  \otimes \limits_{i = n + 1}^{2n} {{\rm H}^{{1 \over 3}\left( {h_i^Bc_i^B + h_i^Ac_i^A + 4} \right)}}P_i^C} \right.,
\end{array}
\end{equation}
 (${{\rm H}^2} = I$, and applying $\rm H$ twice to single-qubit does nothing to it). 
Charlie then announces the position information $i$ ($i \le n$) that satisfies ${P_i}^{{C}^{\prime \prime }}= P_i^{C}$ and ${P_{i+n}}^{{C}^{\prime \prime }} \ne  P_{i+n}^{C}$.

\item  Alice and Bob get ${S_{in}}$ based on the results announced by Calvin.
 \end{enumerate}

It is worth noting that in Step 3 and Step 5, through exclusive OR calculation, the private information of a participant is randomly encoded in the last $n$ qubits, which prevents participants from getting ${S_{diff}}$ and avoids the problem of their internal private information leakage.
\section{Correctness and security analysis}
\subsection{Correctness analysis}
The corresponding results of ${P_i}^{{C}^{\prime \prime }}$ for several paired $c_i^A$ and $c_i^B$ are shown in Table ~\ref{tab:1}. The spectral decomposition form of the H gate can be expressed as
${\rm H} = \left| {{y_1}} \right\rangle \left\langle {{y_1}} \right| - \left| {{y_2}} \right\rangle \left\langle {{y_2}} \right|$,
where $\left| {{y_1}} \right\rangle  = \left( {\matrix{
   {\sqrt 2  - 1}  \cr
   {3 - 2\sqrt 2 }  \cr
 } } \right)$ and $\left| {{y_2}} \right\rangle  = \left( {\matrix{
   {3 - 2\sqrt 2 }  \cr
   {1 - \sqrt 2 }  \cr
 } } \right)$. 
We can calculate ${{\rm H}^{{4 \over 3}}} = \left| {{y_1}} \right\rangle \left\langle {{y_1}} \right| + {\left(  -1  \right)^{{4 \over 3}}}\left| {{y_2}} \right\rangle \left\langle {{y_2}} \right| = I$, and ${{\rm H}^{{5 \over 3}}} = \left| {{y_1}} \right\rangle \left\langle {{y_1}} \right| + {\left(  -1  \right)^{{5 \over 3}}}\left| {{y_2}} \right\rangle \left\langle {{y_2}} \right| =\rm H$.
Therefore, if $c_i^A=c_i^B$ ($i \le n$), Calvin can get ${P_i}^{{C}^{\prime \prime }} =  P_i^{C}$. Otherwise, ${P_i}^{{C}^{\prime \prime }} \ne  P_i^{C} $. 
If  $c_{i+n}^A=c_{i+n}^B=0$, Calvin can get ${P_{i+n}}^{{C}^{\prime \prime }} =  P_{i+n}^{C} $. 
If $c_{i+n}^A=c_{i+n}^B=1$, Calvin can get ${P_{i+n}}^{{C}^{\prime \prime }} \ne  P_{i+n}^{C} $ . 
If $c_{i+n}^A \ne c_{i+n}^B$, ${P_{i+n}}^{{C}^{\prime \prime }}$ and $  P_{i+n}^{C} $ may be equal or not.
So, if ${P_i}^{{C}^{\prime \prime }} =P_i^{C}$  and ${P_{i+n}}^{{C}^{\prime \prime }} \ne  P_{i+n}^{C}$, the $i$-th number is the ${S_{in}}$ element.
\begin{table}
\caption{QPSI protocol results}
\label{tab:1}      
\begin{tabular}{lllll}
\hline\noalign{\smallskip}
 Case &  $c_i^A $ & $c_i^B$ & ${P_i}^{{C}^{\prime \prime }}$($i \le n$)  & ${P_i}^{{C}^{\prime \prime }}$($n < i \le 2n$) \\
\noalign{\smallskip}\hline\noalign{\smallskip}
1 & 0 & 0 & ${{\rm H}^{{4 \over 3}}}P_i^C$ & ${{\rm H}^{{4 \over 3}}}P_i^C$  \\
2 & 0 & 1 & ${{\rm H}^{{5 \over 3}}}P_i^C$ & ${{\rm H}^{{5 \over 3}}}P_i^C$/ ${{\rm H}^{{4 \over 3}}}P_i^C$ \\
3 & 1 & 0 & ${{\rm H}^{{5 \over 3}}}P_i^C$ & ${{\rm H}^{{4 \over 3}}}P_i^C$/ ${{\rm H}^{{5 \over 3}}}P_i^C$ \\
4 & 1 & 1 & $ {{\rm H}^{{2}}}P_i^C$ & ${{\rm H}^{{5 \over 3}}}P_i^C$  \\
\noalign{\smallskip}\hline
\end{tabular}
\end{table}

We use an example to verify the correctness of the protocol.
In this section, insertion of decoy photons and the eavesdropper detection are omitted.
Suppose Alice and Bob have private sets ${S_A} = \left\{ {5,7,17,20} \right\}$ and ${S_B} = \left\{ {7,13,17,35} \right\}$, respectively. The universal set is $U = \left\{ {2,5,7,9,13,17,20,35} \right\}$. They encode sets ${S_A}$ and ${S_B}$ as ${C_A} = \left\{ {0,1,1,0,0,1,1,0} \right\}$ and ${C_B} = \left\{ {0,0,1,0,1,1,0,1} \right\}$, respectively.

In Step 1, Calvin prepares the quantum sequence ${P_{C}} = \left| {101 + 00 -  + +01 + -0 - 1 } \right\rangle $.
In Step 3, Alice prepares ${R_{A}} = \left\{ {3,6,4,2,4,1,9,7,3,6,4,2,4,1,9,7} \right\}$ and ${\rm H_A} = \left\{ {1,1,0,0,1,0,0,1} \right\}$. And she executes the $\rm H$ gates on the ${P_{C}}$, she gets 
\begin{equation}
\label{eq9}
   \begin{array}{rcl}
   {P_{A}} =&& \mathop  \otimes \limits_{i = 1}^8 {{\rm H}^{{1 \over 3}\left( {c_i^A + 2} \right) + r_i^{A}}}P_i^{C1}\left\| {\mathop  \otimes \limits_{i = 9}^{16} {{\rm H}^{{1 \over 3}\left( {h_i^Ac_i^A + 2} \right) + r_i^A}}} \right.\\
   =&& {{\rm H}^{{2 \over 3} + 3}}\left| {\rm{1}} \right\rangle  \otimes {\rm H}{{\rm H}^{{\rm{1 + 6}}}}\left| {\rm{0}} \right\rangle  \otimes {{\rm H}^{{\rm{1 + 4}}}}\left| 1 \right\rangle  \otimes {{\rm H}^{{2 \over 3} + 2}}\left|  +  \right\rangle  \\
   && \otimes {{\rm H}^{{2 \over 3} + 4}}\left| 0 \right\rangle  \otimes {{\rm H}^{{\rm{1 + 1}}}}\left| 0 \right\rangle  \otimes {{\rm H}^{{\rm{1 + 9}}}}\left|  -  \right\rangle  \otimes {{\rm H}^{{2 \over 3} + 7}}\left|  +  \right\rangle \\
   &&\otimes {{\rm H}^{\rm{3}}}\left|  +  \right\rangle  \otimes {{\rm H}^{{5 \over 3}{\rm{ + 5}}}}\left| {\rm{0}} \right\rangle  \otimes {{\rm H}^6}\left| 1 \right\rangle  \otimes {\rm H}\left|  +  \right\rangle   \\
   &&  \otimes {{\rm H}^3}\left|  -  \right\rangle  \otimes {{\rm H}^9}\left| 0 \right\rangle  \otimes {{\rm H}^2}\left|  -  \right\rangle  \otimes {{\rm H}^4}\left| 1 \right\rangle .
        \end{array}
\end{equation}

In Step 5, Bob prepares random numbers ${R_{B}} = \left\{ {7,9,8,5,4,2,3,6,7,9,8,5,4,2,3,6} \right\}$. He calculates ${\rm H_B} = \left\{ {0,0,1,1,0,1,1,0} \right\}$ and executes the $\rm H$ gates on the ${P_{A}}$. He gets:
\begin{equation}
\label{eq10}
   \begin{array}{rcl}
   {P_{B}}=&& \mathop  \otimes \limits_{i = 1}^8 {{\rm H}^{{1 \over 3}\left( {c_i^B + c_i^A + 4} \right) + r_i^B + r_i^A}}P_i^C\left\| {\mathop  \otimes \limits_{i = 9}^{16} {{\rm H}^{{1 \over 3}\left( {h_i^Bc_i^B + h_i^Ac_i^A + 4} \right) + r_i^B + r_i^A}}P_i^C} \right.\\
   =&&{{\rm H}^{{4 \over 3} + 10}}\left| {\rm{1}} \right\rangle  \otimes {{\rm H}^{{5 \over 3} + 15}}\left| {\rm{0}} \right\rangle  \otimes {{\rm H}^{2 + 12}}\left| 1 \right\rangle  \otimes {{\rm H}^{{4 \over 3} + 7}}\left|  +  \right\rangle  \\
   && \otimes {{\rm H}^{{5 \over 3} + 8}}\left| 0 \right\rangle  \otimes {{\rm H}^{2 + 3}}\left| 0 \right\rangle  \otimes {{\rm H}^{{5 \over 3} + 12}}\left|  -  \right\rangle  \otimes {{\rm H}^{{5 \over 3} + 13}}\left|  +  \right\rangle \\
   &&\otimes{{\rm H}^{{4 \over 3} + 4}}\left|  +  \right\rangle  \otimes {{\rm H}^{{5 \over 3} + 8}}\left| {\rm{0}} \right\rangle  \otimes {{\rm H}^{{5 \over 3} + 11}}\left| 1 \right\rangle  \otimes {{\rm H}^{{4 \over 3} + 3}}\left|  +  \right\rangle \\
   && \otimes {{\rm H}^{{4 \over 3} + 7}}\left|  -  \right\rangle  \otimes {{\rm H}^{{5 \over 3} + 12}}\left| 0 \right\rangle  \otimes {{\rm H}^{{4 \over 3} + 8}}\left|  -  \right\rangle  \otimes {{\rm H}^{{4 \over 3} + 13}}\left| 1 \right\rangle .
        \end{array}
\end{equation}

 In Step 7, Alice and Bob calculate $r_i^{C} = r_i^{A} + r_i^{B}$ and send the results to Calvin. Calvin gets:
\begin{equation}
\label{eq11}
   \begin{array}{rcl}
   {P_{C}}^{\prime \prime }=&& \mathop \otimes \limits_{i = 1}^n {{\rm H}^{r_i^{C}}}P_i^{B}\\
   =&&{{\rm H}^{10}}{{\rm H}^{{4 \over 3} + 10}}\left| {\rm{1}} \right\rangle  \otimes {{\rm H}^{15}}{{\rm H}^{{5 \over 3} + 15}}\left| {\rm{0}} \right\rangle  \otimes {{\rm H}^{12}}{{\rm H}^{2 + 12}}\left| 1 \right\rangle  \otimes {{\rm H}^7}{{\rm H}^{{4 \over 3} + 7}}\left|  +  \right\rangle  \\
   &&   \otimes {{\rm H}^8}{{\rm H}^{{5 \over 3} + 8}}\left| 0 \right\rangle  \otimes {{\rm H}^3}{{\rm H}^{2 + 3}}\left| 0 \right\rangle  \otimes {{\rm H}^{12}}{{\rm H}^{{5 \over 3} + 12}}\left| {\rm{ - }} \right\rangle  \otimes {{\rm H}^{13}}{{\rm H}^{{5 \over 3} + 13}}\left|  +  \right\rangle \\
   &&\otimes{{\rm H}^4}{{\rm H}^{{4 \over 3} + 4}}\left|  +  \right\rangle  \otimes {{\rm H}^8}{{\rm H}^{{5 \over 3} + 8}}\left| {\rm{0}} \right\rangle  \otimes {{\rm H}^{11}}{{\rm H}^{{5 \over 3} + 11}}\left| 1 \right\rangle  \otimes {{\rm H}^3}{{\rm H}^{{4 \over 3} + 3}}\left|  +  \right\rangle  \\

   && \otimes {{\rm H}^7}{{\rm H}^{{4 \over 3} + 7}}\left|  -  \right\rangle  \otimes {{\rm H}^{12}}{{\rm H}^{{5 \over 3} + 12}}\left| 0 \right\rangle  \otimes {{\rm H}^8}{{\rm H}^{{4 \over 3} + 8}}\left|  -  \right\rangle  \otimes {{\rm H}^{13}}{{\rm H}^{{4 \over 3} + 13}}\left| 1 \right\rangle  \\
   =&&{{\rm H}^{{4 \over 3}}}\left| {\rm{1}} \right\rangle  \otimes {{\rm H}^{{5 \over 3}}}\left| {\rm{0}} \right\rangle  \otimes \left| 1 \right\rangle  \otimes {{\rm H}^{{4 \over 3}}}\left|  +  \right\rangle \\
   &&\otimes {{\rm H}^{{5 \over 3}}}\left| 0 \right\rangle  \otimes \left| 0 \right\rangle  \otimes {{\rm H}^{{5 \over 3}}}\left| {\rm{ - }} \right\rangle  \otimes {{\rm H}^{{5 \over 3}}}\left|  +  \right\rangle \\
   &&\otimes{{\rm H}^{{4 \over 3}}}\left|  +  \right\rangle  \otimes {{\rm H}^{{5 \over 3}}}\left| {\rm{0}} \right\rangle  \otimes {{\rm H}^{{5 \over 3}}}\left| 1 \right\rangle  \otimes {{\rm H}^{{4 \over 3}}}\left|  +  \right\rangle \\

   && \otimes {{\rm H}^{{4 \over 3}}}\left|  -  \right\rangle  \otimes {{\rm H}^{{5 \over 3}}}\left| 0 \right\rangle  \otimes {{\rm H}^{{4 \over 3}}}\left|  -  \right\rangle  \otimes {{\rm H}^{{4 \over 3}}}\left| 1 \right\rangle  \\
   =&&\left| {\rm{1}} \right\rangle  \otimes {\rm H}\left| {\rm{0}} \right\rangle  \otimes \left| 1 \right\rangle  \otimes \left|  +  \right\rangle  \otimes {\rm H}\left| 0 \right\rangle  \otimes \left| 0 \right\rangle  \otimes {\rm H}\left| {\rm{ - }} \right\rangle  \otimes {\rm H}\left|  +  \right\rangle \\
   &&\otimes\left|  +  \right\rangle  \otimes {\rm H}\left| {\rm{0}} \right\rangle  \otimes {\rm H}\left| 1 \right\rangle  \otimes \left|  +  \right\rangle  \otimes \left|  -  \right\rangle  \otimes {\rm H}\left| 0 \right\rangle  \otimes \left|  -  \right\rangle  \otimes \left| 1 \right\rangle .
        \end{array}
\end{equation}
He obtains ${P_1}^{{C}^{\prime \prime }} = P_1^{C}$, ${P_3}^{{C}^{\prime \prime }} = P_3^{C}$, ${P_4}^{{C}^{\prime \prime }} = P_4^{C}$, ${P_6}^{{C}^{\prime \prime }} = P_6^{C}$,
$P_{10}^{C} \ne {P_{10}}^{{C}^{\prime \prime }}$, 
$P_{11}^{C} \ne {P_{11}}^{{C}^{\prime \prime }}$,
$P_{14}^{C} \ne {P_{14}}^{{C}^{\prime \prime }}$. Calvin gets the 3rd and 6th position to meet the conditions and announces it to Alice and Bob.
In the end, Alice and Bob get ${S_{in}} = \left\{ {7,17} \right\}$.

\subsection{Security analysis}
In this section, we conduct security analysis from two aspects: participant attack and outside attack.

\subsubsection{Participant attack}
Participant attack\cite{RefG07,RefS13} means that dishonest participants try to steal other's information during the calculation process, which will lead to the leakage of private information inside the participants. 
Ensuring the internal information security of participants is one of the important criteria for evaluating the security of the protocol.
\paragraph{Case 1:} Alice tries to steal Bob's private information.

During the protocol process, Alice can only receive the quantum sequence ${P_C} = \mathop \otimes \limits_{i = 1}^n P_i^C$ sent by Calvin, and $P_i^C$ is randomly determined by Calvin, so Alice cannot obtain Bob's private information in this case.

At the end of the protocol, Alice gets the information of $c_i^A = c_i^B = 1$, and then gets ${S_{in}}$. In other cases, she does not know whether the element satisfies $c_i^A = c_i^B = 0$ or $c_i^A \ne c_i^B$, so she cannot deduce $c_i^B = 1$ and cannot obtain Bob's private information outside ${S_{in}}$.
\paragraph{Case 2:}Bob tries to steal Alice's private information.

Bob gets sequence ${P_{A}}$ (as shown in Eq.~(\ref{eq3})) after receiving the decoy photon information published by Alice.
If dishonest Bob wants to use the measure-resend attack to get Alice's private information, he prepares the ancillary qubits $\left| 0 \right\rangle _B^n$ and entangles Alice's quantum sequence $P_i^A$ that has been received with the ancillary qubits. He wants to get Alice's private information $c_i^A$ by measuring the ancillary qubits sequence. After receiving  $P_i^{A}$, Bob uses the unitary operator ${\tilde U_{AB}}$ to operate on $P_i^A$ and $\left| 0 \right\rangle _B^n$. According to Ref.\cite{RefL2019}, Bob's attack process is as follows:
\begin{equation}
{\tilde U_{AB}}P_i^{A}\left| 0 \right\rangle _B^n = \sqrt \eta  P_i^{A}\left| {u\left( {P_i^{A}} \right)} \right\rangle  + \sqrt {1 - \eta } {\left| {v\left( {P_i^{A}} \right)} \right\rangle _{AB}} \;,
\end{equation}
where $P_i^{A}\left| {u\left( {P_i^{A}} \right)} \right\rangle $ and ${\left| {v\left( {P_i^{A}} \right)} \right\rangle _{AB}}$ are orthogonal vectors.
 \begin{equation}
{\left\langle {v\left( {P_i^A} \right)} \right|_{AB}}P_i^{A}\left| {u\left( {P_i^{A}} \right)} \right\rangle  = 0.
\end{equation}
In order to successfully pass the eavesdropping detection stage, Eve's operation will not change the state of the original photon, so $\eta $ should be equal to 1. He can get:
\begin{equation}
   \begin{array}{rcl}
   {{\tilde U}_{AB}}P_i^{A}\left| 0 \right\rangle _B^n &&= P_i^{A}\left| {u\left( {P_i^{A}} \right)} \right\rangle \\

    &&= {{\rm H}^{{1 \over 3}\left( {h_i^Ac_i^A + 2} \right) + r_i^{A}}}P_i^C\left| {u\left( {P_i^{A}} \right)} \right\rangle \\

   && =\left\{
\begin{array}{rcl}
{{{\rm H}^{{2 \over 3} + r_i^{A}}}P_i^{C}\left| {u\left( {P_i^{A}} \right)} \right\rangle ,if c_i^A = 0}\\
{{{\rm H}^{1+r_i^{A}}}P_i^{C}\left| {u\left( {P_i^{A}} \right)} \right\rangle ,if c_i^A = 1}\;.
\end{array} \right.
        \end{array}
\end{equation}
Bob cannot extract the global phase information from the partial qubits of the entangled quantum, so he cannot obtain any information about Alice.

At the end of the protocol, Bob gets the information of $c_i^A = c_i^B = 1$ and gets ${S_{in}}$. 
In other cases, similar to Alice, he does not know whether the element satisfies $c_i^A = c_i^B = 0$ or $c_i^A \ne c_i^B$, so he cannot deduce $c_i^A = 1$ and cannot obtain Alice's private information outside ${S_{in}}$.
\paragraph{Case 3:} Calvin tries to steal Alice's and Bob's private information.

After Calvin receives the decoy photon information published by Bob, he gets the sequence ${P_{B}}$.
\begin{equation}
\label{eq113}
{P_{B}} = \mathop  \otimes \limits_{i = 1}^n {{\rm H}^{{1 \over 3}\left( {c_i^B + 2} \right)}}{{\rm H}^{r_i^B}}P_i^A\left\| {\mathop  \otimes \limits_{i = n + 1}^{2n} {{\rm H}^{{1 \over 3}\left( {h_i^Bc_i^B + 2} \right)}}{{\rm H}^{r_i^B}}P_i^A} \right..
\end{equation}
Calvin receives $r_i^C$ and operates on sequence ${P_{B}}$ to get the new sequence ${P_{C}}^{\prime \prime }$.
\begin{equation}
\label{eq14}
{P_C}^{\prime \prime } = \mathop  \otimes \limits_{i = 1}^n {{\rm H}^{{1 \over 3}\left( {c_i^B + c_i^A + 4} \right)}}P_i^C\left\| {\mathop  \otimes \limits_{i = n + 1}^{2n} {{\rm H}^{{1 \over 3}\left( {h_i^Bc_i^B + h_i^Ac_i^A + 4} \right)}}P_i^C} \right..
\end{equation}
Because the last $n$ qubits information is randomly determined by Alice and Bob, Charlie can only infer the private information of Alice and Bob based on the first $n$ qubits. If ${P_i}^{{C}^{\prime \prime }} \ne P_i^{C}$ ($i \le n$) , Calvin can deduce that the $i$-th position element is in the private set of Alice or Bob (i.e. $c_i^A = 1$ or $ c_i^B = 1$), but he is not sure which one.
Therefore, Calvin cannot obtain all the private information of Alice or Bob, and the protocol satisfies privacy.

\subsubsection{Outside attack}
At present, the common quantum channel attacks include intercept-measuring-retransmission attack, Trojan horse attack \cite{RefF06}, man-in-the-middle attack \cite{RefP09}, invisible photon attack \cite{RefK05}, etc. In the communication between the two parties, the third-party attacker Eve is the object with strong attack ability, and its eavesdropping technology is only limited by the basic principles of quantum mechanics. The decoy photon eavesdropping detection method \cite{RefL05,RefM05} is an important method to detect whether there is eavesdropping in the quantum communication process. Its safety has been proved in refs\cite{RefY13}.

In our protocol, the external attacker Eve can attack the quantum channel during the transmission of the photon sequence between Calvin, Alice and Bob. In these steps, the participants insert some particles into the quantum sequence in the form of decoy states. Take the information transfer between Calvin and Alice in Step 1 to Step 2 as an example. When Calvin sends the quantum sequence inserted into the decoy photon to Alice, if Eve launches an attack, he will inevitably change the original sequence after stealing the information. After receiving the quantum sequence, Alice can judge whether there is a malicious attack based on the position information of the decoy photon and the measurement base information given by Calvin.

\section{Conclusion}
In this paper, we point out the problem of private information leakage between participants in the NQPSI protocol \cite{RefL21}. To solve this problem, we propose an improved QPSI protocol based on $\rm H$ gates.
We use the more feasible $\rm H$ gates to replace the original quantum Fourier transform, which may reduce the difficulty of the protocol implementation. 
Through the exclusive OR calculation on the last n qubits, this scheme ensures that participants cannot get ${S_{diff}}$, which prevents them from getting the private information of the others outside ${S_{in}}$.
It is worth noting that if the third party is malicious or there is an error in the protocol process, the participant will get wrong results without knowing it. In this regard, attention should be paid to the verifiability of QPSC and the integrity of third parties. At present, the verifiable blind quantum computing framework based on the idea of delegating private computations \cite{RefLC18} has attracted much attention. We believe that learning from this idea to solve the PSC problem is a feasible way.

\begin{acknowledgements}
The authors would like to thank the anonymous reviewers and editors for their comments that improved the quality of this paper. This work is supported by the National Natural Science Foundation of China (62071240, 61802002), and the Priority Academic Program Development of Jiangsu Higher Education Institutions (PAPD).
\end{acknowledgements}
\section*{Declarations}

\textbf{Conflict of interest} The authors declare that they have no conflict of interest.\\
\textbf{Ethical statement} Articles do not rely on clinical trials.\\
\textbf{Human and animal participants} All submitted manuscripts containing research which does not involve human participants and/or animal
experimentation.

\end{document}